\documentclass{article}

\usepackage[main, final]{neurips_2026}

\usepackage[utf8]{inputenc} 
\usepackage[T1]{fontenc}    
\usepackage{hyperref}       
\usepackage{url}            
\usepackage{booktabs}       
\usepackage{amsfonts}       
\usepackage{nicefrac}       
\usepackage{microtype}      
\usepackage{xcolor}         
\usepackage{float}          
\usepackage{wrapfig}        

\newcommand{\name}[1]{\textsc{MultiLoop}}

\usepackage{booktabs}
\usepackage{tabularx}
\usepackage{pifont}

\usepackage{array}
\newcolumntype{C}[1]{>{\centering\arraybackslash}p{#1}}

\usepackage{graphicx}
\usepackage{subcaption}

\usepackage{algorithm}
\usepackage[noend]{algpseudocode}
\usepackage{amsmath,amssymb}

\usepackage{listings}
\usepackage{tcolorbox}
\tcbuselibrary{listings, skins}

\usepackage{xltabular}

\usepackage{tikz}
\usetikzlibrary{positioning}
\usepackage{wrapfig}

\definecolor{googleblue}{HTML}{4285F4}
\definecolor{googlelightgray}{HTML}{F8F9FA}

\definecolor{diffincl}{rgb}{0.1, 0.5, 0.1}
\definecolor{diffrem}{rgb}{0.7, 0.1, 0.1}
\lstdefinelanguage{diff}{
  morecomment=[f][\color{diffincl}]{+},
  morecomment=[f][\color{diffrem}]{-},
}

\newtcblisting{diffbox}[1]{
  colback=googlelightgray,     
  colframe=googleblue,         
  colbacktitle=googleblue,     
  coltitle=white,              
  title={\texttt{#1}},
  listing only,
  listing options={
    language=diff,
    basicstyle=\ttfamily\tiny,
    breaklines=true,
    breakatwhitespace=false
  },
  left=2mm, right=2mm, top=2mm, bottom=2mm,
  arc=4pt,                     
  boxrule=1pt                  
}

\title{\name{}: Evolving Multi-Agent Harnesses for Complex Information-Seeking Tasks}

\author{%
  Daniela Gottesman$^{1}$\thanks{Work done as an intern at Google Research.} \quad
  Gal Yona$^{2}$ \quad
  Ori Ram$^{2}$ \quad
  Zorik Gekhman$^{2}$ \\
  \bfseries
  Jonathan Herzig$^{2}$ \quad
  Mor Geva$^{1}$ \quad
  Roee Aharoni$^{2}$ \\
  $^{1}$Tel Aviv University \quad $^{2}$Google Research \\
  \texttt{gottesman3@tau.mail.ac.il}
}

\begin{document}

\maketitle
\begin{abstract}

The performance of LLM-based agents depends not only on the underlying model but also on the harness: the prompts, tools, and orchestration logic around it. Recent works automate harness optimization with coding agents that analyze execution traces, diagnose failures, and edit the harness code; however, those focused on single-agent systems. Multi-agent harnesses are harder to optimize as fixing them requires changing how agents interact in addition to how each behaves. We study whether automated harness optimization can diagnose and repair failures in multi-agent systems through a case study on complex information-seeking tasks from DeepSearchQA \citep{deepsearchqa}. We apply \name{}, a simple optimization loop in which a coding agent iteratively analyzes, evaluates and edits the harness. With only 10 training samples, the evolved harness improves accuracy by 4.6\% over the base harness and by 3.2\% over the best single-agent system. However, a taxonomy of failure modes before and after optimization shows that the orchestration failures that characterize multi-agent systems largely persist. Our findings motivate harness optimization methods that encourage structural changes to how agents delegate tasks and share information.

\end{abstract}
\section{Introduction}

Agentic behavior in large language models (LLMs) is typically enabled by a ``harness’’: the surrounding infrastructure that defines prompts, access to tools, and orchestration flow. As agentic systems grow more complex and produce longer trajectories, manually identifying harness failures has become impractical, motivating recent efforts to automate both detection and mitigation. These works automate harness optimization using coding agents to iteratively analyze execution traces, diagnose failures, and improve the harness code \citep{lee2026meta, lou2026autoharnessimprovingllmagents, zhang2026self}. Crucially, for these optimizers to accurately diagnose failures and propose updates, they require full visibility into historical code changes, execution traces, and evaluation results. This optimization process, however, has been primarily studied for single-agent systems.

Multi-agent harnesses, where an orchestrator agent delegates work to subagents, are a common and effective design for long-horizon tasks \citep{zhu2024redeltoolkitllmpoweredrecursive, grand2025selfsteering, schroeder-etal-2025-thread, zhang2026recursive, multiagentsystem, alzubi2026romarecursiveopenmetaagent}.
However, shifting from single- to multi-agent harnesses complicates harness optimization. For example, reading the full rollouts of many subagents can overwhelm the optimizer’s context window \citep{zhang2025which}. More fundamentally, failures may also cascade from poor orchestration, such as delegating overly broad tasks, parallelizing dependent ones, or sharing too little context between agents, which leads to redundant work and long verification loops \citep{cemri2026why, wang2026agentdropoutv2}. Unlike in single-agent systems, fixing these would require changing not only how individual agents behave but also how they interact. To our knowledge, no prior work has examined whether automated harness optimization can diagnose and repair such failures.

In this work, we take a first step toward answering this question through a case study on automatically evolving a multi-agent harness for complex information-seeking tasks, using Google's DeepSearchQA benchmark \citep{deepsearchqa}. This is a natural setting as it produces long, complex trajectories involving reasoning, tool use, and sandboxed environments. In addition, these tasks have been shown to benefit from multi-agent systems with subagent delegation which alleviate long-context issues and enable decomposition of sub-tasks \citep{zhang2026recursive}.

Rather than proposing a new optimization strategy, we apply an existing paradigm to a multi-agent harness to understand which failures it resolves and which remain. Following \citet[]{lee2026meta, zhang2026self}, we employ \name{}, a simple automated optimization loop where a coding agent iteratively improves a base multi-agent harness. In each iteration, an Optimizer agent evaluates the latest rollouts on a small training set of examples to identify weaknesses and failure modes in the current harness. With access to the history of past scores and prior harness edits, the agent consolidates recurring issues among different rollouts and executes targeted code edits to yield a new, optimized harness. Finally, the proposed harness is evaluated, and the loop repeats. 

We find that harness optimization does transfer to the multi-agent setting: the evolved harness improves accuracy by 4.61\% over the base harness and outperforms the best single-agent system by 3.19\%. However, these improvements come from prompt refinements, such as stricter answer formatting and instructions for processing tool outputs programmatically, and from the addition of a new tool, rather than from structural orchestration changes that improve coordination between agents. As a result, the orchestration failures characteristic of multi-agent systems largely persist. We support this with a taxonomy and qualitative evaluation of the failure modes before and after optimization. Ultimately, our findings motivate future work on harness optimization that encourages structural changes to multi-agent harnesses, such as redesigning how the orchestrator delegates tasks or what information agents pass to one another.
\section{Harness Optimization}
\label{sec:method}
\begin{algorithm}[H]
\scriptsize
\caption{\textsc{MultiLoop}: automated multi-agent harness optimization}
\label{alg:multiloop}
\begin{algorithmic}[1]
\State \textbf{Input:} training set $\mathcal{X}$, base harness $H_0$, iterations $N$, LLM $M$
\State $A_{0}, R_0 \gets \Call{Evaluate}{H_0, M, \mathcal{X}}$ \Comment{Evaluate base harness on training set to get initial accuracy $A_0$ and rollouts $R_0$}
\State $\mathcal{L} \gets \{(H_0, A_0, R_0)\}$ \Comment{Stores all harness versions, and their training set accuracy and rollouts}
\For{$t = 1 \ldots N$}
    \State $H_t \gets{} \Call{Optimize}{H_{t-1}, A_{t-1}, R_{t-1}, \mathcal{L}}$ \Comment{Optimizer proposes a new harness based on previous evaluation and history}
    \State $A_{t}, R_t \gets \Call{Evaluate}{H_t, M, \mathcal{X}}$ \Comment{Evaluate the new harness on the training set}
    \State $\mathcal{L} \gets \mathcal{L} \cup \{(H_t, A_t, R_t)\}$ \Comment{Add the new harness, accuracy, and rollouts to the history}
\EndFor
\State \Return $\arg\max_{H_t} \{ A_t \mid (H_t, A_t, R_t) \in \mathcal{L} \}$
\end{algorithmic}
\end{algorithm}

This section presents our method for autonomous optimization of multi-agent harnesses. Algorithm~\ref{alg:multiloop} implements \name{} as an agentic workflow with two subroutines: \textsc{Evaluate} and \textsc{Optimize}. We initialize \name{} with a simple harness in which an orchestrator delegates tasks to subagents, each executing tool calls in its own context (see Appendix~\ref{sec:visual}, Figure~\ref{fig:multiloop-acdl} for a visualization). 

\paragraph{\textsc{Evaluate}}
Given a harness $H_t$, \textsc{Evaluate} runs the current agentic system independently on every training example in $\mathcal{X}$, producing a rollout per example, collectively denoted $R_t$. To make these rollouts consumable by the Optimizer, each is programmatically converted into a markdown file that distinguishes orchestrator steps from subagent steps and renders every step's model thoughts, tool calls, tool outputs, and generated answer. Finally, accuracy $A_t$ is computed over the training set (explained in \S~\ref{sec:evaluation}) and appended to a running history of harness versions with their accuracies and rollouts, $\mathcal{L}$, allowing the optimization loop to rollback from a bad harness version.

\paragraph{\textsc{Optimize}}
The Optimizer is a coding agent that reads the evaluation outputs in $\mathcal{L}$ and proposes code edits to the harness. It is implemented with the Antigravity coding-agent SDK, which configures an agent with a system prompt (found in Appendix~\ref{app:prompts}), and a set of tools for filesystem navigation (file viewing, directory listing, and grep-based text search), code editing (file editing and creation), and shell execution (for checking code compilation and performing version-control operations).

\section{Experiments}
\label{sec:experiments}
\paragraph {Experimental Setup} We evaluate \name{} on DeepSearchQA \citep{deepsearchqa}, a recent 900-prompt benchmark for evaluating agents on difficult multi-step information-seeking tasks. As iterative harness optimization is costly, the training set includes 10 challenging examples that all of the unoptimized agent systems in Table~\ref{tab:meta_agent_results} could not solve, and the test set is the remaining 890 examples. This is motivated by recent evidence that meaningful optimization signal can be extracted from a small number of well-chosen examples \citep{wang2025reinforcement}, and significant experimental costs as described in Appendix ~\ref{sec:price_breakdown}. Across all experiments, we used Gemini-3-Flash as the model in the evaluated harness, and Gemini-3.1-Pro with the Antigravity harness as the Optimizer \citep{geminiteam2025gemini3}. 

\begin{wraptable}{r}{0.5\textwidth} 
\centering
\scriptsize
\setlength{\tabcolsep}{2pt}
\caption{Fully correct accuracy (FC) on 890 DeepSearchQA test questions.}
\label{tab:meta_agent_results}
\begin{tabular}{@{}lcc@{}}
\toprule
\textbf{Agent Architecture} & \textbf{Multi-agent} & \textbf{FC} \\
\midrule
\multicolumn{3}{@{}l}{\textit{Without harness optimization}} \\
Search Agent             & no & $58.35 \pm 2.29$ \\
Search + Other Tools (SPOT)   & no & $68.69 \pm 2.16$ \\
Orchestrator w/ SPOT subagents & yes & $67.27 \pm 2.18$ \\
\midrule
\multicolumn{3}{@{}l}{\textit{With full harness optimization}} \\
SPOT                           & no & $65.54 \pm 2.21$ \\
Orchestrator w/ SPOT subagents & yes & $\mathbf{71.88} \pm 2.09$ \\
\midrule
\multicolumn{3}{@{}l}{\textit{With prompt only opt.}} \\
Orchestrator w/ SPOT subagents  & yes & $64.62 \pm 2.22$ \\
\bottomrule
\end{tabular}
\end{wraptable}

\paragraph{Baselines}
To evaluate our approach, we first compare against three baselines without any optimization. (1) Search Agent, a multi-step agent which only has access to a Google search tool; (2) Search Plus Other Tools (SPOT) Agent, which adds tools for website scraping, PDF scraping, and Python execution; and (3) Orchestrator with SPOT subagents, our baseline multi-agent harness. We then run three optimization experiments. First is (4) SPOT + harness optimization, showing the effect of optimization on a single-agent harness, (5) is Orchestrator w/ SPOT subagents + harness optimization which optimizes the multi-agent harness, and finally (6) Orchestrator w/ SPOT subagents + prompt optimization, an ablation experiment which only optimizes the system instructions, injected messages, and function declarations. All optimization loops are run for 10 iterations. 

\paragraph{Evaluation}
\label{sec:evaluation}
We measure the percentage of fully correct (FC) responses, defined as an exact semantic match to the ground-truth entity set (i.e., identifying all correct answers with zero incorrect ones) using Gemini-3-Flash \citep{geminiteam2025gemini3} and the DeepSearchQA \citep{deepsearchqa} grader prompt. Results show the average FC and 95\% confidence interval averaged across two runs on the 890-question test set, which is disjoint from the training set.

\paragraph{Results}
Table \ref{tab:meta_agent_results} shows that adding additional tools on top of the search tool improves performance, with the SPOT agent outperforming the Search Agent by 10.34\%. Adding the naive multi-agent harness on top of that does not show immediate benefits, with the unoptimized Orchestrator with SPOT subagents underperforming the single SPOT agent by 1.42\%. Applying \name{} improves multi-agent performance by 4.61\% to 71.88\%, which is the best result we obtained on the test set. On the single-agent harness, optimization degraded the performance by 3.15\%. For the ablation allowing only prompt optimization to the multi-agent harness, we see a performance decrease of 2.65\%.

\section{Analysis}

We now unpack the factors behind these results: why optimization degrades single-agent performance yet improves multi-agent performance, and which harness changes drove that improvement and how they reshaped the distribution of failures over the course of optimization.

\paragraph{Optimization harms single agent performance} We find that the single-agent degradation stems from an optimization step that added 30K-character truncation for tool outputs, which did not generalize to the test set. This suggests that a larger training set could be useful, however we kept the training set minimal at 10 samples due to compute constraints (see Appendix \ref{sec:price_breakdown}).

\paragraph{Analyzing the multi-agent optimization results} 

\begin{wrapfigure}{r}{0.45\textwidth} 
    \centering
    \includegraphics[width=\linewidth]{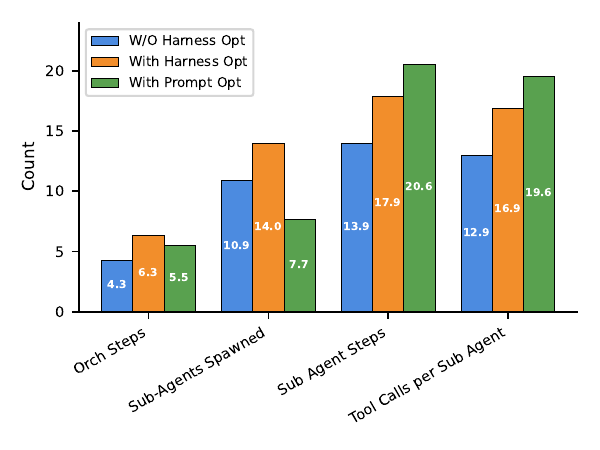}
    \caption{Comparison of trajectory statistics for Orchestrator with SPOT subagents.}
    \label{fig:tool_call_distr}
\end{wrapfigure}

Figure~\ref{fig:tool_call_distr} compares the three multi-agent harnesses: unoptimized (Blue), full harness optimization (Orange), and prompt-only optimization (Green). Statistics are averaged over two evaluation runs on the 890-question test set.

Full harness optimization makes the system do more work overall: the orchestrator takes more steps (6.3 vs. 4.3), spawns more subagents (14.0 vs. 10.9), and each subagent runs longer (17.9 vs. 13.9 steps), yielding 250 subagent steps and 236 subagent tool calls per question versus 152 and 141 for the unoptimized harness. Prompt-only optimization instead spawns the fewest subagents (7.7 per question), but each one issues the longest trajectories: 20.6 steps and 19.6 tool calls on average, compared with 13.9 steps and 12.9 tool calls without optimization and 17.9 and 16.9 with full optimization, and 46\% of prompt-only subagents hit the 30-step limit compared with 19\% and 35\%, respectively. As a result, the prompt-only system does roughly the same total subagent work as the unoptimized one (158 vs. 152 steps per question) while being less accurate. This suggests that prompt-only optimization overfit on the train set, and pushed individual subagents into longer, less effective trajectories rather than better delegation, pointing at the value of allowing the optimizer to make structural changes. Because full optimization increases total inference, part of the gain may reflect additional compute rather than better harness design---a compute-matched comparison \citep{tran2026singleagentllmsoutperformmultiagent} is left for future work.

\paragraph{Taxonomy and Distribution of Failure Modes} 

We manually analyzed 11 trajectories from the Orchestrator with SPOT subagents run where the final output shifted from fully incorrect to fully correct (or vice versa) post-optimization. By cross-referencing observations with the Trajectory Audit Categories proposed by \citet{li2026sgr}, we established 7 categories, fully detailed in Appendix~\ref{sec:failure_taxonomy}: 
\textbf{Bad Orchestration} (poor decisions about when, how, and how many subagents to dispatch); \textbf{Ineffective Tool Usage} (mechanical failures in tool selection and execution); \textbf{Retrieval Scope Drift} (losing track of key task constraints mid-task); \textbf{Criterion Mismatch} (applying an incorrect decision rule, field definition, or time window to an otherwise relevant source); \textbf{Tool Output Misreading} (misreading a value, label, or field from the correct source); \textbf{Retrieval Dependency Errors} (failing to connect correct intermediate results across sources or subagents); and \textbf{Final Answer Composition} (verbose final answers when a concise one is required).

\begin{wrapfigure}{r}{0.5\textwidth} 
    \centering
    \includegraphics[width=\linewidth]{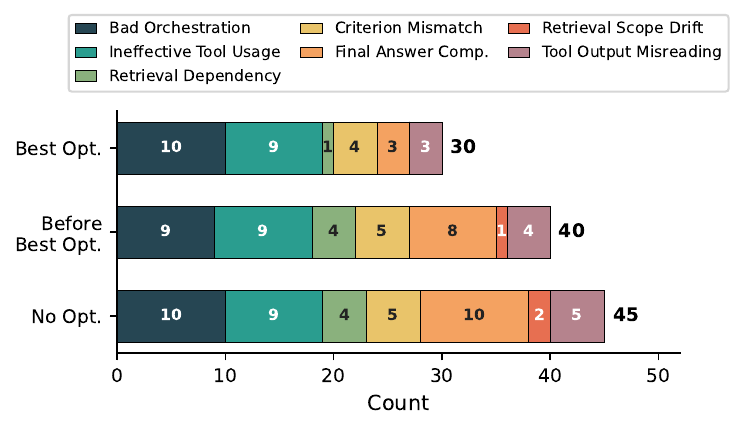}
    \caption{Distribution of failure modes.}
    \label{fig:distribution_failures}
\end{wrapfigure}

To better understand how \name{} resolves specific errors, we analyze the rollouts over the training set from the Orchestrator w/ SPOT subagents setting, by counting the number of examples that contain at least one instance of a given error category. As illustrated in Fig.~\ref{fig:distribution_failures}, the baseline model (No Opt.) has the highest overall frequency of errors, dominated by bad orchestration, ineffective tool usage, and final answer composition failures. The intermediate stage (Before Best Opt.) shows incremental progress. The best optimization configuration (Best Opt.) reduces overall failures, primarily by resolving final answer composition errors, retrieval dependencies, and retrieval scope drift. Orchestration and tool usage errors were not alleviated with harness optimization, indicating that these may require model improvements or more dedicated optimization logic.

\paragraph{Harness Changes Made by \name{}}
As outlined in Appendix~\ref{sec:appendix_architecture}, Tables~\ref{tab:error_mitigation_1} and \ref{tab:error_mitigation_2}, the adjustments to the Orchestrator w/ SPOT subagents harness were primarily at the prompt-level (e.g., system instructions, function declarations, and error message propagation to agents). The only architectural change was the addition of a new tool (\texttt{download\_file}) for retrieving files from the internet. The majority of the modifications focused on refining subagent tool usage by clarifying use-cases. To address orchestration deficits, the orchestrator's system prompt was updated to recommend query decomposition, parallel delegation across subagents---mirroring the delegation principle in Anthropic's Claude multi-agent system \citep{multiagentsystem} and the supervisor/worker pattern in LangGraph \citep{langgraph2024}---and specific tool use behavior for subagents. Further, to mitigate tool output misreading, the subagents were given explicit directives on how to parse raw string outputs into JSON objects for programmatic analysis---similar to what \citet{schick2023toolformer} and \citet{hong2024metagpt} solve with typed structured schemas---and included an example of composing multiple tool calls within the python sandbox. Finally, errors in final answer composition were addressed through instructions that encouraged concise responses. The jump to the Best Opt. score of 71.88\% stemmed specifically from refinements to the download tool and stricter final-answer formatting instructions. Consistent with \citet{ursekar2026vero}, aside from introducing the download tool, the optimizer avoided deeper structural changes (e.g., modifying the response schema between subagents and the orchestrator) in favor of surface-level prompt and formatting adjustments.

\paragraph{Analyzing Changes in System Behavior}
Building on our analysis of the harness changes, we also look at how the agent's behavior changed after harness optimization. 9 questions improved from fully incorrect to fully correct, while 2 regressed. Appendix~\ref{sec:behavior_before_after}, Tables~\ref{tab:behavioral_changes_part1} and \ref{tab:behavioral_changes_part2} showcase behavioral differences observed between the rollouts before and after optimization: improvements were mainly driven by programmatic processing of outputs via the \texttt{download\_file} tool (6 examples) and better, sequential orchestration of dependent tasks (3 examples), while the 2 degraded examples failed because the orchestrator became trapped in long verification loops.

\section{Related Work}

\paragraph{Issues with Multi-Agent Harnesses}
Multi-agent harnesses are a common design for long-horizon tasks because they give each subtask a fresh context window and let independent subtasks run in parallel \citep{multiagentsystem, zhu2024redeltoolkitllmpoweredrecursive, schroeder-etal-2025-thread, zhang2026recursive, alzubi2026romarecursiveopenmetaagent}. However, these systems require careful multi-agent coordination and communication. \citet{kim2026towards} find that coordination helps on decomposable tasks, hurts on sequential ones, and propagates errors across agents unless a central coordinator catches them. \citet{tran2026singleagentllmsoutperformmultiagent} show that under fixed compute, a single agent matches or beats multi-agent systems, due to information loss in hand-offs. \citet{ou2025analyzinginformationsharingcoordination} find that structured information sharing and a reflective orchestrator can improve performance of multi-agent systems. Failure taxonomies place most multi-agent errors in task specification and inter-agent misalignment \citep{cemri2026why}, and even attributing a failure to the responsible agent and step is hard \citep{zhang2025which}. Our taxonomy in Appendix~\ref{sec:failure_taxonomy}, is consistent with these findings: bad orchestration dominates our failure distribution, and it is a category that harness optimization struggles to improve.

\paragraph{Automatic Harness Optimization}
Earlier methods optimize prompts from LLM-generated feedback \citep{yang2024large, yuksekgonul2024textgradautomaticdifferentiationtext, agrawal2026gepa} or search over agentic workflows represented as code \citep{hu2025automated, zhang2025aflow}. Recent work treats the full harness code as the optimization variable and uses a coding agent to propose and implement harness edits in an iterative loop \citep{lee2026meta, zhang2026self, ursekar2026vero, zhang2026darwingodelmachineopenended, lou2026autoharnessimprovingllmagents, lin2026agenticharnessengineeringobservabilitydriven, nie2026tthetesttimeharnessevolution, chen2026harnessxcomposableadaptiveevolvable, karten2026continualharnessonlineadaptation}. Two limitations of this paradigm are directly relevant to our study. First, evolved harnesses may overfit the training set, with gains that shrink or vanish on held-out tasks \citep{wang2026rethinking, xia2026rrsiregularizedrecursiveselfimprovement}. Second, optimizers favor surface-level edits such as prompt rewrites over structural changes \citep{ursekar2026vero, xia2026rrsiregularizedrecursiveselfimprovement}. A concurrent work, RRSI \citep{xia2026rrsiregularizedrecursiveselfimprovement} mitigates both by regularizing proposal and selection.
\section{Conclusions and Future Work} 
We present a case study in which we apply \name{}, a simple harness-optimization loop, to a multi-agent system for complex information-seeking tasks. On DeepSearchQA \citep{deepsearchqa}, \name{} improves the multi-agent harness, but the gains come from prompt refinements and a new tool rather than from changes to how agents delegate work or share information, and the orchestration failures that characterize multi-agent systems largely persist. Future work includes larger-scale experiments, compute-matched comparisons against single-agent baselines, and optimization methods that explicitly encourage structural changes to orchestration, such as redesigning task decomposition or the information passed between agents.

\begin{ack}
We thank Tomer Levinboim for helpful discussions on early versions of this project.
\end{ack}

\bibliographystyle{plainnat}
\bibliography{custom}

\appendix

\section{Appendix}

\section{Limitations}
\label{sec:limitations}
For technical reasons, we were restricted to use Gemini models and Antigravity during the course of this work. It would be interesting to extend such experiments to additional models and harnesses. Another limitation is that we only used DeepSearchQA, and extending our evaluation to additional benchmarks would be appealing. Other knobs we would have liked to explore altering with more time and compute available is the number of optimization iterations, and the number and composition of the samples in the training set.

\section{Visualization of Multi-Agent System}
\label{sec:visual}

\begin{figure}[H]
    \centering
    \begin{subfigure}[t]{0.48\textwidth}
      \vspace{0pt}
      \includegraphics[width=\textwidth]{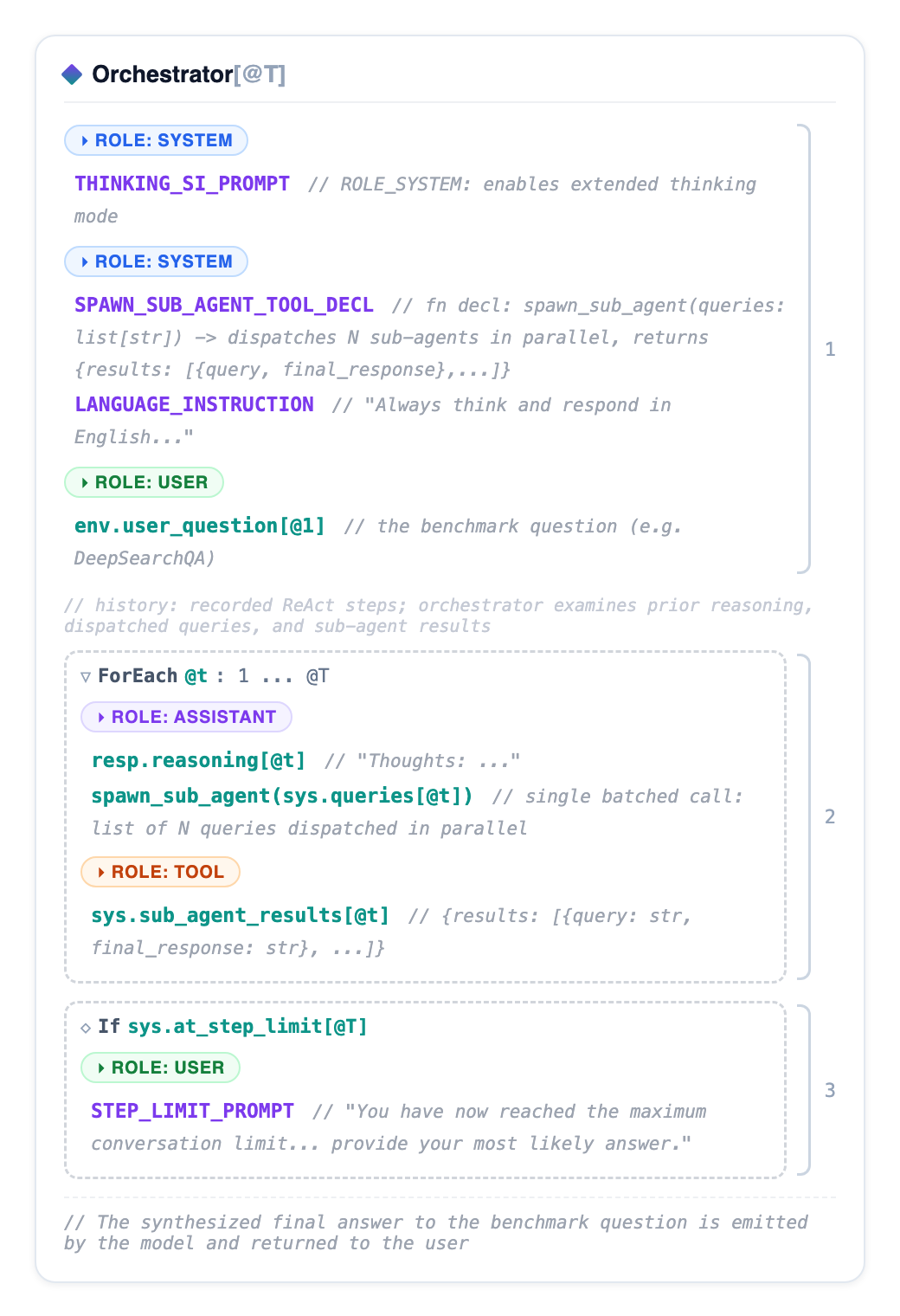}
    \end{subfigure}
    \hfill
    \begin{subfigure}[t]{0.51\textwidth}
      \vspace{0pt}
      \includegraphics[width=\textwidth]{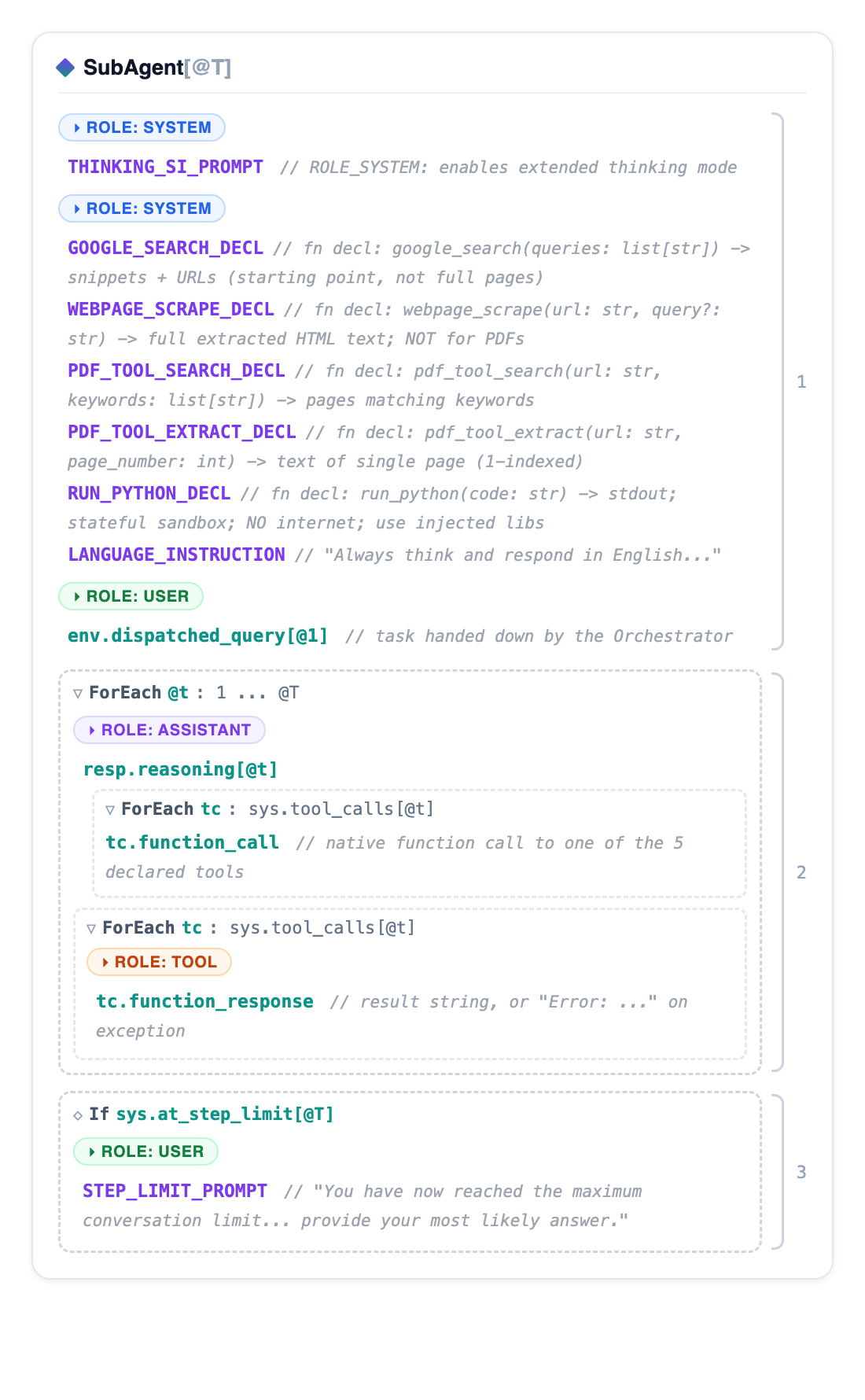}
    \end{subfigure}
    \caption{The initial \name{} orchestrator (left) and its subagent (right).}
  \label{fig:multiloop-acdl}
 
\end{figure}

Figure \ref{fig:multiloop-acdl} illustrates the initial multi-agent harness, showing how the orchestrator delegates work to subagents via \texttt{spawn\_sub\_agent} while only receiving each subagent's final answer back into its context.

\section{Optimizer Prompts}
\label{app:prompts}

\subsection*{System Prompt — Orchestration Mode}

\begin{quote}\ttfamily\small
You are a harness engineer improving a multi-agent QA system. You have
full permission to modify the codebase---edit, create, or delete files
to improve the system. Analyze the rollouts, identify failure patterns,
and make targeted code changes. Commit and upload your changes when done.
\end{quote}

\subsection*{System Prompt for the Optimizer— Prompt-Only Mode}

\begin{quote}\ttfamily\small
You are a prompt engineer improving a multi-agent QA system. You may
ONLY edit \{prompts\_file\}. Analyze the rollouts, identify failure
patterns, and make targeted prompt changes. Commit and upload your
changes when done.
\end{quote}

\subsection*{User Prompt for the Optimizer (both modes)}

\begin{quote}\ttfamily\small
\#\#\# Score History\\
\{table of all prior rounds: version $\to$ \%~correct, F1, P, R\}\\[4pt]
\#\#\# Rollouts\\
\{path to rollout dir\}. Each example subdirectory contains
summary.json (prompt, expected answer, agent response) and rollout.md
(step-by-step tool calls and outputs).\\[4pt]
\#\#\# Instructions\\
1.~Read rollouts to understand failures.\\
2.~Identify failure patterns.\\
3.~Make targeted changes.\\
4.~Verify the code builds.\\
5.~Commit and upload.
\end{quote}

\section{Price Breakdown of Experimental Setup}
\label{sec:price_breakdown}

We limited the number of optimization steps, and training examples due to budget constraints. A single evaluation over the test set (890 questions × 2 runs) costs about \$7.3K for the unoptimized harness, \$12.9K for the fully optimized harness, and \$9.5K for the prompt-only harness (mean \$4.25, \$7.15, and \$5.25 per question), and each evaluation of the 10-question training set costs \$104 and \$213 for the unoptimized and optimized harnesses (mean \$10 and \$21 per question, since these are the hardest examples).

\section{Failure Mode Taxonomy}
\label{sec:failure_taxonomy}

\textbf{Bad Orchestration}
The orchestrator makes poor decisions about when, how, and how many subagents
to dispatch. This includes spawning parallel subtasks that have logical
dependencies, sending overly broad or keyword-style prompts to subagents,
spawning redundant subagents, over-parallelizing subagents, and falling into
verification loops instead of outputting the final answer.
\textbf{Ineffective Tool Usage}
Agents fail at the mechanical level of tool selection and execution, such as
picking the wrong tool for the content type (e.g. using the browse tool to
examine pdfs), being blocked by sandbox limitations, not following through on
stated tool intentions, and using the Python tool to simply call existing tools
without further analysis of their results.
\textbf{Retrieval Scope Drift}
The agent loses track of key constraints mid-task, or fails to validate that
all required entities satisfy those constraints before producing a final answer.
\textbf{Criterion Mismatch}
The agent reaches a relevant source but applies an incorrect decision rule,
field definition, aggregation level, or time window. The resulting answers
often contain locally correct values bound to the wrong criterion, like using
data from the wrong year.
\textbf{Tool Output Misreading}
The agent reaches the correct page or data source but misreads a local value,
label, or field. Some common examples are misinterpreting tool outputs like
flattened tables, confusing similar-looking fields or rows, or struggling to
extract data from interactive charts and paginated results that require dynamic
interaction.
\textbf{Retrieval Dependency Errors}
The agent obtains correct intermediate results from individual sources but
fails to properly connect them, such as failing to resolve conflicting data
across sources or failing to join datasets from separate subagents into a
comparative table.
\textbf{Final Answer Composition}
Agents produce verbose final answers with markdown tables, caveats, and
paragraph explanations when a specific, concise answer is required.

\section{System Architecture Updates}
\label{sec:appendix_architecture}
Table \ref{tab:error_mitigation_1} and Table \ref{tab:error_mitigation_2} summarize the concrete harness and prompt changes implemented to address each identified failure category, from orchestration instructions to tool-output parsing guidance.

\renewcommand{\arraystretch}{1.5} 

\begin{table}[H] 
\footnotesize
\centering
\renewcommand{\arraystretch}{1.5}
\begin{tabularx}{\textwidth}{@{} l X @{}}
\toprule
\textbf{Error Category} & \textbf{Implemented Harness Change} \\
\midrule

\textbf{Bad Orchestration} & 
\textit{Updated the orchestrator's system instruction to direct how to manage subagents:} 
{\setlength{\leftskip}{1em} \ttfamily\raggedright\color{teal!70!black} You are the Orchestrator. Your job is to break down complex queries into smaller, independent sub-queries and spawn subagents to solve them. When a task involves comparing data across multiple entities (e.g., countries, years, boroughs), ALWAYS dispatch separate subagent queries for each entity so they can be researched in parallel. Be highly specific in your queries to the subagents so they know exactly what data to find.} \\
\addlinespace[0.5em]

\textbf{Ineffective Tool Usage} & 
\textit{Modified the orchestrator's system instruction to instruct subagents to use specific tools for specific tasks:} 
{\setlength{\leftskip}{1em} \ttfamily\raggedright\color{teal!70!black} If the task involves analyzing datasets or tabular data, explicitly instruct the subagents to use their `run\_python` tool to download and process the data.
}
\textit{Modified error messages in failed tool executions:} 
{\setlength{\leftskip}{1em} \ttfamily\raggedright\color{teal!70!black} f'ERROR: Failed to extract page \{page\_number\} from \{url\} \par
Reason: \{e\} \par
The PDF may be unavailable, require authentication, or the URL may point to an HTML page instead of a PDF. Try using the `browse` tool on this URL.'} 

\textit{Updated the subagent's system instruction on how to use tools:} 
{\setlength{\leftskip}{1em} \ttfamily\raggedright\color{teal!70!black} CRITICAL: When searching for specific statistics, numbers, or tabular data, Google Search snippets usually do NOT contain the exact answer. Instead of iteratively guessing numbers in your search queries, you MUST use the `browse` tool to read the full page of the most promising URLs returned by `google\_search`. For PDF documents, use the `pdf\_tool\_search` and `pdf\_tool\_extract` tools. For datasets, CSVs, or complex data analysis, use the `run\_python` tool to download the data with `download\_lib.download\_file()` and analyze it using `pandas`.}

\textit{Added a hint in the function declaration of the google\_search tool:} 
{\setlength{\leftskip}{1em} \ttfamily\raggedright\color{teal!70!black} 
If the exact answer is not in the snippets, do NOT guess numbers in new queries; instead, use the `browse` tool on the returned URLs.
}

\textit{Added a new tool for downloading online files, like CSVs:} 
{\setlength{\leftskip}{1em} \ttfamily\raggedright\color{teal!70!black}  
def download\_file(url: str, save\_path: str = "downloaded\_file") -> str: \par
\hspace*{1em}"""Downloads a file from the internet to the local sandbox.""" \par
\hspace*{1em}...}
\\
\bottomrule
\end{tabularx}
\caption{Examples of optimization mitigations for failure modes. (Part 1)}
\label{tab:error_mitigation_1}
\end{table}

\begin{table}[H]
\footnotesize
\centering
\renewcommand{\arraystretch}{1.5} 
\begin{tabularx}{\textwidth}{@{} l X @{}}
\toprule
\textbf{Error Category} & \textbf{Implemented Harness Change} \\
\midrule

\textbf{Tool Output Misreading} & 
\textit{Updated the subagent's system instruction to explain the use-cases of different tools:} 
{\setlength{\leftskip}{1em} \ttfamily\raggedright\color{teal!70!black} CRITICAL: When searching for specific statistics, numbers, or tabular data, Google Search snippets usually do NOT contain the exact answer. Instead of iteratively guessing numbers in your search queries, you MUST use the `browse` tool to read the full page of the most promising URLs returned by `google\_search`. For PDF documents, use the `pdf\_tool\_search` and `pdf\_tool\_extract` tools. For datasets, CSVs, or complex data analysis, use the `run\_python` tool to download the data with `download\_lib.download\_file()` and analyze it using `pandas`.
}

\textit{Updated the function declarations of the tools to instruct the subagents how to use the tools in the sandboxed environment and parse their outputs:} 
{\setlength{\leftskip}{1em} \ttfamily\raggedright\color{teal!70!black} You MUST use the provided tools. NOTE: ALL injected tools return RAW STRINGS (str). If the returned string is in JSON format, you MUST parse it using `json.loads()` before indexing into it as a dictionary or list... \par
str and extract\_page(url: str, page\_number: int) -> str \par
- download\_lib: download\_file(url: str, save\_path: str = "downloaded\_file") -> str \par
Use `download\_lib` to download CSV, JSON, or other data files into your sandbox. DO NOT use requests/urllib, they will fail. \par
Call the tools directly in your python code. For example: \par
\par
\hspace*{1em}import json \par
\hspace*{1em}res = json.loads(google\_search\_lib.search(queries=["q1"])) \par
\hspace*{1em}download\_lib.download\_file( \par 
\hspace*{2em}"https://example.com/data.csv", "data.csv") \par
\hspace*{1em}import pandas as pd \par
\hspace*{1em}df = pd.read\_csv("data.csv") \par
\hspace*{1em}print(df.head()) 
}
\\
\addlinespace[0.5em]

\textbf{Final Answer Comp.} & 
\textit{Added a strict guardrail in the orchestrator's system instruction:} 
{\setlength{\leftskip}{1em} \ttfamily\raggedright\color{teal!70!black} CRITICAL INSTRUCTION FOR FINAL RESPONSE: \par
When you have gathered the required information, you MUST output ONLY the exact requested answer as your final text. \par
ABSOLUTELY NO CONVERSATIONAL FILLER. Do not use phrases like "Based on...", "According to...", or "The answer is...". If the answer is a list of items, provide a simple comma-separated list of those items and NOTHING ELSE. Any extra text, explanations, or context will cause the test to FAIL.} \\

\bottomrule
\end{tabularx}
\caption{Examples of optimization mitigations for failure modes. (Part 2)}
\label{tab:error_mitigation_2}
\end{table}

\section{Behavioral Changes: Before vs. After Examples}
\label{sec:behavior_before_after}
Table \ref{tab:behavioral_changes_part1} and Table \ref{tab:behavioral_changes_part2} contrast pre- and post-optimization rollout behavior across representative questions, demonstrating measurable shifts such as adopting programmatic data processing, resolving sequential dependencies correctly, and browsing full pages instead of relying on search snippets.

\begin{table}[H]
\centering
\caption{Examples of optimization mitigations for failure modes. (Part 1)}
\label{tab:behavioral_changes_part1}
\renewcommand{\arraystretch}{1.5}
\begin{tabularx}{\textwidth}{>{\hsize=0.8\hsize}X >{\hsize=1.0\hsize}X >{\hsize=1.1\hsize}X >{\hsize=1.1\hsize}X}
\toprule
\textbf{Behavioral Change} & \textbf{Example \& Question} & \textbf{``Before'' Rollout Quote / Action} & \textbf{``After'' Rollout Quote / Action} \\
\midrule

\textbf{Programmatic Data Processing} \par (Using Python to download and parse data instead of searching) &
\par \textbf{Question:} ``Based on OWID sources for records of the year 2021... which country had the largest child mortality rate also in 2021.'' &
\textbf{Quote:} ``I'll use the Python tool to find the specific values or rankings by searching for `top age-standardized death rates 2021' on the web or looking for the GBD 2021 publication.'' \par \textbf{Action:} The agent improperly attempts to use the Python environment merely to run iterative Google search queries without downloading data, resulting in hallucinated values from snippets. &
\textbf{Quote:} ``Since I can't browse the interactive table results easily via search snippets, I will use Python to search and maybe download if I find a CSV link.'' \par \textbf{Action:} The subagent successfully uses Python to circumvent sandbox limitations, utilizing the injected \texttt{download\_lib} tool to fetch the CSV datasets, parse them with pandas, and extract the real values for 2021. \\
\midrule

\textbf{Sequential Dependency Resolution} \par (Waits for intermediate facts before proceeding) &
\par \textbf{Question:} ``According to Toyota, what is the part number for the oil filter cap assembly for the highest MSRP 2020 Toyota SUV that uses a battery with the same chemistry as the 2021 Toyota Aqua?'' &
\textbf{Action:} The Orchestrator attempted to solve the logically dependent steps at once by blindly guessing the target vehicle in parallel. It dispatched queries like: ``oil filter cap assembly part number for 2020 Toyota Land Cruiser'' and ``oil filter cap assembly part number for 2020 Toyota Highlander Hybrid'' before even identifying the required battery chemistry. &
\textbf{Quote:} ``What is the battery chemistry of the 2021 Toyota Aqua?'' and ``Which 2020 Toyota SUV models use a battery with the same chemistry as the 2021 Toyota Aqua?'' \par \textbf{Action:} The Orchestrator avoids guessing and correctly assigns subagents sequentially to identify the Bipolar NiMH battery before executing a targeted search for the Highlander Hybrid's oil filter cap. \\

\bottomrule
\multicolumn{4}{r}{\textit{Continued on next page...}} \\
\end{tabularx}
\end{table}

\begin{table}[H]
\centering
\caption{Examples of optimization mitigations for failure modes. (Part 2)}
\label{tab:behavioral_changes_part2}
\renewcommand{\arraystretch}{1.5}
\begin{tabularx}{\textwidth}{>{\hsize=0.8\hsize}X >{\hsize=1.0\hsize}X >{\hsize=1.1\hsize}X >{\hsize=1.1\hsize}X}
\toprule
\textbf{Behavioral Change} & \textbf{Example \& Question} & \textbf{``Before'' Rollout Quote / Action} & \textbf{``After'' Rollout Quote / Action} \\
\midrule

\textbf{Page Browsing} \par (Opening URLs to read full text instead of relying on snippets) &
\par \textbf{Question:} (Implied from context) Calculate the horsepower-to-cargo-volume ratio for specific Honda vehicles based on Kelley Blue Book (KBB) data. &
\textbf{Quote:} ``I'll present the calculation using Maximum Cargo Volume as it's the standard for general comparison.'' \par \textbf{Action:} The agent relied entirely on the \texttt{google\_search} snippets. Without reading the full context of the spec sheet, it incorrectly chose the minivan's maximum cargo volume (144.9 cu. ft. with seats removed) instead of the standard trunk space, skewing the math. &
\textbf{Quote:} ``I will use Google Search to find the specific kbb.com page for the 2023 Honda HR-V and then use the browse tool to extract the required specifications.'' \par \textbf{Action:} Guided by a new instruction to use the browse tool, the agent establishes the correct detailed intention. This rigorous approach causes it to accurately extract the standard cargo volume behind the third row (``32.8 cu. ft.''), leading to the correct mathematical conclusion. \\
\midrule

\textbf{Over-Delegation \& Exhaustive Verification (Degradation)} \par (Over-thinking and getting trapped in loops) &
\par \textbf{Question:} (Implied from context) Find the most recently seen species among those assessed as `Extinct' (EX) on the IUCN Red List that occurred in Spain. &
\textbf{Action:} The subagent correctly zeroes in on multiple candidate species and executes Python looping scripts to cross-reference their last-seen dates via google search, rapidly deducing that the freshwater snail \textit{Islamia ateni} was the most recent. &
\textbf{Quote:} ``Are there any others? I'll use google search within python to find the specific list of EX insects in Spain.'' \par \textbf{Action:} Even though the Orchestrator established the correct answers internally, it became paranoid about proving a negative. It redundantly spawned multiple subagents to hunt for obscure species, trapping the system in an endless verification loop. \\

\bottomrule
\end{tabularx}
\end{table}

\newpage

\end{document}